\documentclass[pra,superscriptaddress,showpacs,showkeys,notitlepage]{revtex4-1}
\usepackage{amsmath}
\usepackage{color}

\begin{document}
\title{Self-consistent kinetic equations for $e^-e^+\gamma$-plasma \\ generated from vacuum by strong electric field}

\author{S.A. Smolyansky}
\affiliation{Saratov State University, 410026 Saratov, Russia}
\affiliation{Laboratory of Quantum Theory of Intense Fields, Tomsk State University, Lenin Prospekt 36, 634050, Tomsk, Russia}
\author{A.D. Panferov}
\affiliation{Saratov State University, 410026 Saratov, Russia}
\author{S.O. Pirogov}
\affiliation{Saratov State University, 410026 Saratov, Russia}
\author{A.M. Fedotov}\email{Email: am\_fedotov@mail.ru}
\affiliation{National Research Nuclear University (MEPhI), 115409 Moscow, Russia}
\affiliation{Laboratory of Quantum Theory of Intense Fields, Tomsk State University, Lenin Prospekt 36, 634050, Tomsk, Russia}

\pacs{12.20.Ds, 52.25.Dg}
\keywords{$e^+e^-$ pair production, strong electric fields, quantum radiation, kinetic equations}

\begin{abstract}
We present a self-consistent kinetic description of an electron-positron-photon ($e^-e^+\gamma$) plasma generated from vacuum in a focal spot of counterpropagating laser pulses. We rely on a model of purely time-dependent external electric field, but properly take into account quantum radiation from plasma, as well as the semiclassical internal (plasma) field.  To achieve this goal we derive a system of coupled kinetic equations for the electron, positron, and photon plasma species supplemented with Maxwell equation for the internal electric field. Quantum radiation is included systematically by taking into account the pair creation/annihilation and radiation channels of the BBGKY chain in quasiparticle representation, which we truncate at second order of perturbation theory. An important application of our results would be consistent consideration of photon emission and cascade production beyond the locally constant field approximation, including the depletion of the driving laser field and a possible scenario of cascade saturation due to thermalization.
\end{abstract}
\maketitle
\section{Introduction}
\label{intro}
Due to the recent advances in femtosecond laser technology it becomes possible to generate extremely strong electromagnetic fields right in a laboratory \cite{Yanovsky2004,Yanovsky2008}.  A few laser facilities worldwide being under  different stages of construction, installation or configuration (e.g., Apollon 10PW \cite{Apollon10}, 
ELI Beamlines \cite{ELI-beams}, ELI-NP \cite{ELI-NP}, SULF \cite{SULF} and the 4-PW facility at GIST \cite{GIST}) aim at the new intensity frontier $10^{23}$W/cm$^2$, where one expects strong signatures of radiation reaction \cite{Zhidkov2002,Pukhov2010,Tamburini2010}. Furthermore, the announced even more ambitious projects now in an R\&D/preparatory phase (e.g., OPAL \cite{OPAL}, SEL \cite{SEL}, Gekko EXA \cite{GEKKO_EXA}, ELI main pillar \cite{ELI_Whitebook}, XCELS \cite{XCELS}) may be capable for attaining the next laser intensity level $\sim 10^{24}-10^{25}$W/cm$^2$. Previously, laser-induced QED processes could be studied only by combining a powerful laser with an electron collider (optionally, in an all-optical setup) \cite{SLAC1,SLAC2,PRX1,PRX2}. However, being faced with the probable attainability of such extreme fields in quite a near future and in order to accurately identify its experimental signatures, one needs proper theoretical tools to simulate laser-plasma/vacuum interactions in a regime when the QED processes induced by a strong external field  \cite{Ritus,diPiazzaReview} may occur massively or even dominate the laser plasma dynamics \cite{bellkirk,Ref8,Ref9,Gonoskov2}. 

As of now, most of the known simulations of this challenging interaction regime have been produced by treating the QED events (hard photon emission, pair photoproduction) within the locally constant field approximation (LCFA) and including them into a PIC code via Monte Carlo event generators \cite
{Ref9,Ridgers,Gonoskov1,Grismayer}. However, recently a number of authors have questioned the validity of LCFA suggesting to reconsider its applicability conditions \cite{Harvey,Dinu,Raicher,diPiazza_LCFA,Ilderton2018}. Further progress in studies of macroscopic signatures of the nonlinear strong field QED effects  [massive electron-positron plasma (EPP) production from vacuum, radiation from the created EPP, cascade processes] requires onward systematic development of the kinetic approaches based on both the quasiparticle \cite{Ref1,Ref2} and Wigner \cite{Ref3} representation. Here we make a step towards this direction by presenting an extension of the previously derived kinetic model based on quasiparticle representation (see, e.g., \cite{Ref1,Ref2} and the reviews \cite{Ref4,Ref5}) now taking additionally into account the quantum radiation effects, namely, such interactions of the created EPP with the photon reservoir as hard photon emission/absorption and pair photoproduction/one-photon annihilation. The resulting coupled kinetic equations (KEs) consist of (i) the KE describing EPP production and dynamics under the combined action of a linearly polarized time-dependent external electric field and the quantized electromagnetic field; (ii) the hard photons KE controlling quantum radiation by EPP. Furthermore, these KEs are supplemented with (iii) the Maxwell equation governing the inner plasma field (backreaction). 

The plan of the paper looks as follows: we consider the operator equations in quasiparticle representation in Sec.~\ref{sec-2}, then discuss the main steps of derivation of the kinetic equations and present the results in Sec.~\ref{sec-3}. Finally, a short summary is given in Sec.~\ref{sec-4}.

\section{Notations and preliminaries}
\label{sec-2}
Characteristic property of the problem is the nonperturbative character of spontaneous EPP production from vacuum. Its description is especially simple in a spatially homogeneous time-dependent external electric field 
in the framework of the quasiparticle representation \cite{Ref2}. Though such a background is, strictly speaking, not a solution of Maxwell equations in vacuum, there are still good reasons to believe that it can be used to consider spontaneous EPP production from vacuum in an antinode of a standing wave \cite{Schutzhold}, e.g. created by collision of counterpropagating optical laser pulses \cite{Popov}. While keeping for generality the polarization of the external field arbitrary in the current section, we proceed for simplicity with the case of linearly polarized external field in the rest of the paper. Thus, the vector potential of the total acting semiclassical electric field in the Hamiltonian gauge is $\textbf{A}(t)=\textbf{A}_{ex}(t)+\textbf{A}_{in}(t)$, where $\textbf{A}_{ex}(t)$ and $\textbf{A}_{in}(t)$ correspond to the external and the inner (plasma) fields, respectively. On this background the emission and absorption of hard photons are considered in the framework of perturbation theory with respect to the small parameter $E_q/E_c\ll 1$, where $E_q$ is the characteristic strength field amplitude of a quantized field and $E_c=m^2/e$ is the Schwinger critical field. No restriction except transversality is imposed on  polarization of the fluctuating quantized field with the vector potential $\textbf{A}_q(x)$. Being on shell, the latter is indispensably spatially inhomogeneous.

In the quasiparticle representation the total Hamiltonian can be splitted as follows:
\begin{equation}\label{eq1}
H(t)=H_{qp}(t)+H_s(t)+H_{pol}(t)+H_{ph}+H_{int}(t),
\end{equation}
where the diagonal part $H_{qp}(t)$ corresponds to the EPP quasiparticle subsystem in the total semiclassical field $\textbf{A}(t)$:
\begin{equation}\label{eq2}
H_{qp}(t)=\sum_{\textbf{p}}\omega(\textbf{p},t)[a^{+}_{\alpha}(\textbf{p},t)a_{\alpha}(\textbf{p},t) + b^{+}_{\alpha}(\textbf{p},t)b_{\alpha}(\textbf{p},t)],
\end{equation}
where $\omega(\textbf{p},t)=\sqrt{m^2 + \textbf{P}^2}$ is quasienergy and $\textbf{P}=\textbf{p}-e\textbf{A}(t)$ is quasimomentum.
Here and below summing over twice occurring spin and polarization indices is always implied. The next two parts of the Hamiltonian (\ref{eq1}),
\begin{equation}\label{eq3}
H_s(t)=-i\sum_{\textbf{p}}[a^{+}_{\alpha}(\textbf{p},t)U^{(1)}_{\alpha\beta}(\textbf{p},t)a_{\beta}(\textbf{p},t)-b_{\alpha}(-\textbf{p},t)U^{(1)}_{\alpha\beta}(\textbf{p},t)b^{+}_{\beta}(-\textbf{p},t)],
\end{equation}
and
\begin{equation}\label{3.1}
H_{pol}(t)=-i\sum_{\textbf{p}}[a^{+}_{\alpha}(\textbf{p},t)U^{(2)}_{\alpha\beta}(\textbf{p},t)b^{+}_{\beta}(-\textbf{p},t)-b_{\alpha}(-\textbf{p},t)U^{(2)}_{\alpha\beta}(\textbf{p},t)a_{\beta}(\textbf{p},t)]
\end{equation}
describe the spin precession and the spontaneous pair creation from vacuum along with their reverse annihilation, respectively.
The traceless matrices $U^{(1)}$ and $U^{(2)}$ are given by \cite{Ref6,Ref10,Ref11,Ref12}
\begin{equation}\label{eq9}
U^{(1)}(\textbf{p},t)=i\,\omega\, C(\textbf{p})[\textbf{P}\textbf{E}]\mbox{\boldmath$\sigma$}, \quad
U^{(2)}(\textbf{p},t)=C(\textbf{p})[\textbf{P}(\textbf{P}\textbf{E})-\textbf{E}\,\omega\,\omega_+]\mbox{\boldmath$\sigma$}.
\end{equation}
Here, $\textbf{E}(t)=-\dot{\textbf{A}}(t)$ is the strength of the total electric field, $C(\textbf{p})=e/(2\,\omega^2\,\omega_+)$, $\omega_+=\omega+m$ and $\sigma_i$ are the Pauli matrices.

The quantized radiation field in a plane wave basis and its interaction with the EPP subsystem are described by the Hamiltonians ($k=|\textbf{k}|$)
\begin{equation}\label{eq4}
H_{ph}=\sum_{k} k A_r^{(+)}(\textbf{k},t)A_r^{(-)}(\textbf{k},t),
\end{equation}
and
\begin{eqnarray}\label{eq5}
H_{int}(t)&&=\frac{e}{\sqrt{V}} \sum_{\textbf{p}_1,\textbf{p}_2,\textbf{k}} \frac{1}{\sqrt{2k}} \delta_{\textbf{p}_1 - \textbf{p}_2 + \textbf{k},0} : \biggl\{[\overline{u}u]^r_{\beta \alpha}(\textbf{p}_1,\textbf{p}_2,\textbf{k};t) a^+_\alpha(\textbf{p}_1,t) a^{}_\beta(\textbf{p}_2,t)\notag\\
&&+[\overline{u}\upsilon]^r_{\beta \alpha}(\textbf{p}_1,\textbf{p}_2,\textbf{k};t) a^+_\alpha(\textbf{p}_1,t) b^+_\beta(-\textbf{p}_2,t)+  [\overline{\upsilon}u]^r_{\beta \alpha}(\textbf{p}_1,\textbf{p}_2,\textbf{k};t) b_\alpha(-\textbf{p}_1,t) a^{}_\beta(\textbf{p}_2,t)   \\
&&+[\overline{\upsilon}\upsilon]^r_{\beta \alpha}(\textbf{p}_1,\textbf{p}_2,\textbf{k};t) b_\alpha(-\textbf{p}_1,t) b^+_\beta(-\textbf{p}_2,t)\biggr\} \notag
A_r(\textbf{k},t): ,
\end{eqnarray}
where $V$ is volume of the system and $A_r(\textbf{k},t)=A_r^{(+)}(\textbf{k},t)+A_r^{(-)}(\textbf{k},t)$.

The quasiparticle spinor basis that diagonalizes the Hamiltonian (\ref{eq2}) can be found in the explicit form \cite{Ref6}
\begin{align}\label{eq6}
u^+_1(\textbf{p},t)=B(\textbf{p}) [\omega_+, 0, P^3, P_-], \notag\\
u^+_2(\textbf{p},t)=B(\textbf{p}) [0, \omega_+, P_+, -P^3], \\
\upsilon^+_1(-\textbf{p},t)=B(\textbf{p}) [-P^3, -P_-, \omega_+, 0],\notag \\
\upsilon^+_2(-\textbf{p},t)=B(\textbf{p}) [-P_+, P^3, 0, \omega_+],\notag
\end{align}
where $B(\textbf{p})=(2\omega\omega_+)^{-1/2}$ and $P_\pm=P^1\pm iP^2$. The convolutions (vertex functions) of the spinors $\xi_{\alpha}$ and $\eta_{\alpha}$ from the set (\ref{eq6}) are defined by
\begin{equation}\label{eq7}
[\overline{\xi} \eta]^{r}_{\beta\alpha}(\textbf{p}_1,\textbf{p}_2,\textbf{k};t)=\overline{\xi}_{\alpha}(\textbf{p}_1,t)\gamma^{\mu}\eta_{\beta}(\textbf{p}_2,t)e^{r}_{\mu}(\textbf{k}),
\end{equation}
where $e^{r}_{\mu}(\textbf{k})~(r=1,2)$ are the photon unit polarization vectors. 

The Heisenberg equations of motion with account for the photon subsystem read
\begin{eqnarray}\label{eq8}
\dot{a}_{\alpha}(\mathbf{p},t) &&= -i\omega(\mathbf{p},t)a_{\alpha}(\mathbf{p},t)-U^{(1)}_{\alpha\beta}(%
\mathbf{p},t)a_{\beta}(\mathbf{p},t) -U^{(2)}_{\alpha\beta}(\mathbf{p},t)b^{+}_{\beta} (-\mathbf{p},t) \notag \\
&&- ie\frac{(2\pi)^{3/2}}{V} \sum_{\textbf{p}_1,\textbf{k}}\;  \frac{1}{\sqrt{2k}} \delta_{\mathbf{p}-%
\mathbf{p}_1 +\mathbf{k},0} \; \Bigl\{a_{\beta}(\mathbf{p}_1,t)[\bar{u}u]^{r}_{\beta\alpha}(\mathbf{p},%
\mathbf{p}_1 ,\mathbf{k};t) + b^{+}_{\beta}(-\mathbf{p}_1,t)[\bar{u}v]^{r}_{\beta\alpha}(\mathbf{p},\mathbf{p}_1,\mathbf{k} ;t) %
\Bigr\}A_r (\mathbf{k},t)~,  \notag \\
\dot{b}_{\alpha}(-\mathbf{p},t)&&= -i\omega(\mathbf{p},t)b_{\alpha}(-\mathbf{p},t)+b_{\beta}(-\mathbf{p%
},t)U^{(1)}_{\beta\alpha}(\mathbf{p},t) +a^{+}_{\beta} (\mathbf{p},t)U^{(2)}_{\beta\alpha}(\mathbf{p},t)\notag \\
&&+ ie\frac{(2\pi)^{3/2}}{V} \sum_{\textbf{p}_1,\textbf{k}}\; \frac{1}{\sqrt{2k}}\delta_{\mathbf{p}%
_1 -\mathbf{p}+\mathbf{k},0} \; \Bigl\{[\bar{u}v]^{r}_{\alpha\beta}(\mathbf{p}_1,\mathbf{p},%
\mathbf{k};t) a^{+}_{\beta} (\mathbf{p}_1,t)+ [\bar{v}v]^{r}_{\alpha\beta}(\mathbf{p}_1,\mathbf{p},\mathbf{k};t) b_{\beta}(-\mathbf{p}_1,t)%
\Bigr\}A_r (\mathbf{k},t)~,  \notag \\
\dot{A}^{(\pm)}_r (\mathbf{k},t) &&= \pm ikA^{(\pm)}_r (\mathbf{k},t)\pm
\frac{ie}{\sqrt{2k}}\frac{(2\pi)^{3}}{V} \sum_{\textbf{p}_1,\textbf{p}_2}\; \delta_{\mathbf{p}_1
-\mathbf{p}_2 \mp\mathbf{k},0} \Bigl\{a^{+}_{\beta}(\mathbf{p}_1,t) [\bar{u}u]^{r}_{\alpha\beta}(\mathbf{p}_1 ,\mathbf{p}_2,%
\mp\mathbf{k};t)a_{\alpha}(\mathbf{p}_2 ,t) \notag\\ &&+a^{+}_{\beta}(\mathbf{p}_1,t) [\bar{u}v]^{r}_{\alpha\beta}(\mathbf{p%
}_1 ,\mathbf{p}_2 ,\mp\mathbf{k};t)b^{+}_{\alpha}(-\mathbf{p}_2 ,t) +b_{\beta}(-\mathbf{p}_1,t)[\bar{v}u]^{r}_{\alpha\beta}(\mathbf{p}_1 ,\mathbf{p}_2,\mp\mathbf{k}%
;t) a_{\alpha}(\mathbf{p}_2 ,t) \notag \\ &&+b_{\beta}(-\mathbf{p}_1,t)[\bar{v}v]^{r}_{\alpha\beta}(\mathbf{p}_1 ,%
\mathbf{p}_2,\mp\mathbf{k};t) b^{+}_{\alpha} (-\mathbf{p}_2 ,t)\Bigr\}~. \notag
\end{eqnarray}
In the next section by a standard procedure we pass from these equations to the kinetic ones.

\section{Kinetic equations for $e^-e^+\gamma$-plasma}
\label{sec-3}
A starting point for systematic generalizations is the non-Markovian type KE describing spontaneous EPP creation from vacuum in a linearly polarized time dependent electric field \cite{Ref1,Ref2}, which admits representation either in an integro-differential form
\begin{equation}\label{eq11}
\dot{f}(\textbf{p},t)=I(\textbf{p},t),
\end{equation}
with the source term 
\begin{equation}\label{eq12}
I(\textbf{p},t)=\frac{1}{2}\lambda(\textbf{p},t)\int^{t}dt^{\prime}\lambda(\textbf{p},t^{\prime})[1-2f(\textbf{p},t^{\prime})]\cos\left(2\int^{t}_{t^{\prime}}d\tau\omega(\textbf{p},\tau)\right),
\end{equation}
or in an equivalent form of coupled differential equations:
\begin{equation}\label{eq12.1}
\dot{f}(\textbf{p},t)=\frac{1}{2}\,\lambda(\textbf{p},t)\,u(\textbf{p},t),\quad\dot{u}(\textbf{p},t)=\lambda(\textbf{p},t)\,(1-2\,f(\textbf{p},t)) - 2\,\omega(\textbf{p},t) \,\upsilon(\textbf{p},t),\quad
\dot{\upsilon}(\textbf{p},t)=2\,\omega(\textbf{p},t)\,u(\textbf{p},t),\notag
\end{equation}
where $\lambda(\textbf{p},t)=e\,E(t)\,\epsilon_{\perp}(\textbf{p})/\omega^2(\textbf{p},t)$, $\omega(\textbf{p},t)=\sqrt{ \epsilon_{\perp}^2 +P^2}$ is quasienergy, $\epsilon_{\perp}=\sqrt{P_{\perp}^2+m^2}$ is transversal energy,  and $P=p_{\parallel}-eA(t)$ is longitudinal quasimomentum.

In order to generalize this KE by taking into account interaction with quantized electromagnetic field, we closely follow the steps behind its original derivation \cite{Ref2}. Consider a one-particle correlation function in the quasiparticle representation
\begin{equation}\label{eq13}
f_{\alpha \beta}(\textbf{p},\textbf{p}^{\prime};t)=\langle a^+_{\alpha}(\textbf{p},t) a_{\beta}(\textbf{p}^{\prime},t) \rangle=\langle b^+_{\alpha}(-\textbf{p},t) b_{\beta}(-\textbf{p}^{\prime},t) \rangle,
\end{equation}
where the last equality is a consequence of the electroneutrality condition, which we assume for simplicity (in case of initial excess of electrons or positrons the required modifications are rather obvious).

For simplicity, from now on we assume that both the external and the internal (plasma) fields are linearly polarized, then the spin degrees of freedom are frozen and by the standard assumption of initial correlations fading the correlation function (\ref{eq13}) should be diagonal in both the momentum and the spin space:
\begin{equation}\label{eq14}
f_{\alpha \beta}(\textbf{p},\textbf{p}^{\prime};t)=f(\textbf{p},t)\,\delta_{\alpha \beta}\, \frac{V}{(2\pi)^3}\, \delta_{\textbf{p},\textbf{p}^{\prime}}.
\end{equation}
In the thermodynamic limit Eq. (\ref{eq14}) takes the form
\begin{equation}\label{eq15}
f_{\alpha \beta}(\textbf{p},\textbf{p}^{\prime};t)=f(\textbf{p},t)\,\delta_{\alpha \beta}\,\delta(\textbf{p}-\textbf{p}^{\prime}).
\end{equation}
In an analysis of the Bogoliubov-Born-Kirkwood-Green-Yvon (BBGKY) chains for a $e^-e^+\gamma$ plasma it is convenient to rely on a one-particle two-point correlation functions of the type (\ref{eq13}) possessing the $\delta$-function singularities (\ref{eq15}) in a spatially homogeneous case. Accurate consideration of these singularities is possible in a discrete momentum space.

The first equation of the BBGKY chain in the electron-positron sector reads
\begin{eqnarray}\label{eq16}
\dot{f}(\textbf{p},t)&=&I(\textbf{p},t) + \frac{i\,e}{2\,(2\pi)^3\sqrt{V}}\,\sum_{\textbf{p}_1,\textbf{k}}\frac{1}{\sqrt{2k}}\delta_{\textbf{p}-\textbf{p}_1+\textbf{k},0} \Bigl\{ [\overline{u}u]^{r+}_{\gamma \alpha}(\textbf{p},\textbf{p}_1,\textbf{k};t) \langle a^+_\gamma(\textbf{p}_1,t) a^{}_\alpha(\textbf{p},t) A^{(+)}_r(\textbf{k},t)  \rangle  \notag \\
&+&[\overline{u}\upsilon]^{r+}_{\gamma \alpha}(\textbf{p},\textbf{p}_1,\textbf{k};t) \langle b_\gamma(-\textbf{p}_1,t) a^{}_\alpha(\textbf{p},t) A^{(+)}_r(\textbf{k},t)  \rangle -  [\overline{u}u]^{r}_{\gamma \alpha}(\textbf{p},\textbf{p}_1,\textbf{k};t) \langle a^+_\alpha(\textbf{p},t) a^{}_\gamma(\textbf{p}_1,t) A^{(-)}_r(\textbf{k},t)\rangle \notag\\
&-& [\overline{u}\upsilon]^{r}_{\gamma \alpha}(\textbf{p},\textbf{p}_1,\textbf{k};t) \langle a^+_\alpha(\textbf{p},t) b^{+}_\gamma(-\textbf{p}_1,t) A^{(-)}_r(\textbf{k},t)  \rangle  \Bigr\},\notag
\end{eqnarray}
where the term $I(\textbf{p},t)$ is the same as in Eq. (\ref{eq12}). 

Let us introduce the photon distribution function $F(\textbf{k},t)$ by the relation similar to Eq. (\ref{eq14})
\begin{equation}\label{eq17}
F_{rr^{\prime}}(\textbf{k},\textbf{k}^{\prime},t)=\langle A^{(+)}_r(\textbf{k},t)A^{(-)}_{r^{\prime}}(\textbf{k}^{\prime},t) \rangle=F(\textbf{k},t)\,\delta_{rr^{\prime}}\frac{V}{(2\pi)^3}\,\delta_{\textbf{k},\textbf{k}^{\prime}},
\end{equation}
then the accompanying equation in the photon sector has the form:
\begin{eqnarray}\label{eq18}
\dot{F}(\textbf{k},t)&=&\frac{i\,e}{2\,(2\pi)^3\sqrt{V}}\,\sum_{\textbf{p}_1,\textbf{p}_2}\Bigl\{\frac{1}{\sqrt{2k}}\delta_{\textbf{p}_1-\textbf{p}_2-\textbf{k},0} \notag\\
&& ([\overline{u}u]^{r}_{\beta \alpha}(\textbf{p}_1,\textbf{p}_2,-\textbf{k};t) \langle a^+_\alpha(\textbf{p}_1,t) a^{}_\beta(\textbf{p}_2,t) A^{(-)}_r(\textbf{k},t)  \rangle \notag\\
&+& [\overline{u}\upsilon]^{r}_{\beta \alpha}(\textbf{p}_1,\textbf{p}_2,-\textbf{k};t) \langle a^+_\alpha(\textbf{p}_1,t) b^{+}_\beta(-\textbf{p}_2,t) A^{(-)}_r(\textbf{k},t)  \rangle \notag\\
&+& [\overline{\upsilon}u]^{r}_{\beta \alpha}(\textbf{p}_1,\textbf{p}_2,-\textbf{k};t) \langle b_\alpha(-\textbf{p}_1,t) a^{}_\beta(\textbf{p}_2,t) A^{(-)}_r(\textbf{k},t)  \rangle \notag \\
&+& [\overline{\upsilon}\upsilon]^{r}_{\beta \alpha}(\textbf{p}_1,\textbf{p}_2,-\textbf{k};t) \langle b_\alpha(-\textbf{p}_1,t) b^{+}_\beta(-\textbf{p}_2,t) A^{(-)}_r(\textbf{k},t)  \rangle ) \\
&-& \delta_{\textbf{p}_1-\textbf{p}_2+\textbf{k},0}\,([\overline{u}u]^{r}_{\beta \alpha}(\textbf{p}_1,\textbf{p}_2,\textbf{k};t) \langle a^+_\alpha(\textbf{p}_1,t) a^{}_\beta(\textbf{p}_2,t) A^{(+)}_r(\textbf{k},t)  \rangle \notag\\
&+& [\overline{u}\upsilon]^{r}_{\beta \alpha}(\textbf{p}_1,\textbf{p}_2,\textbf{k};t) \langle a^+_\alpha(\textbf{p}_1,t) b^{+}_\beta(-\textbf{p}_2,t) A^{(+)}_r(\textbf{k},t)  \rangle \notag\\
&+& [\overline{\upsilon}u]^{r}_{\beta \alpha}(\textbf{p}_1,\textbf{p}_2,\textbf{k};t) \langle b_\alpha(-\textbf{p}_1,t) a^{}_\beta(\textbf{p}_2,t) A^{(+)}_r(\textbf{k},t)  \rangle \notag \\
&+& [\overline{\upsilon}\upsilon]^{r}_{\beta \alpha}(\textbf{p}_1,\textbf{p}_2,\textbf{k};t) \langle b_\alpha(-\textbf{p}_1,t) b^{+}_\beta(-\textbf{p}_2,t) A^{(+)}_r(\textbf{k},t)  \rangle) \Bigr\}.\notag
\end{eqnarray}
Each of Eqs. (\ref{eq16}) and (\ref{eq18}) is contributed by the two direct processes (one-photon pair photoproduction and photon emission), and by the two inverse ones (on-photon annihilation and photon absorption) in the right-hand side. The annihilation channel of equation (\ref{eq18}) was previously discussed in Ref.~\cite{Ref6}, where it was applied for a rough estimation of radiation from EPP. Below, we derive a closed system of the coupled KEs for the electron-positron and photon subsystems of the EPP by taking into account all the channels in Eqs. (\ref{eq16}) and (\ref{eq18}) mentioned above.

In order to derive such a closed system of coupled KEs, we make a standard assumption of weak correlation between the single-quasiparticle states due to interaction with quantized electromagnetic field, formally, by first writing the equations of motion for the higher-order correlators arising in the r.h.s. of Eqs. (\ref{eq16}) and (\ref{eq18}), and then closing them by applying in their r.h.s. truncation procedures of the random-phase-approximation (RPA) type \cite{kadanoff}:
\begin{align}\label{eq19}
\langle a^+_{\alpha}(\textbf{p}_1,t) a_{\beta}(\textbf{p}_2,t)A^{(-)}_r(\textbf{k},t)A^{(+)}_{r^{\prime}}(\textbf{k}^{\prime},t) \rangle \approx 
\langle a^+_{\alpha}(\textbf{p}_1,t) a_{\beta}(\textbf{p}_2,t) \rangle \langle A^{(-)}_r(\textbf{k},t)A^{(+)}_{r^{\prime}}(\textbf{k}^{\prime},t) \rangle \\\approx
f(\textbf{p}_1,t)\delta_{\alpha\beta}\delta(\textbf{p}_1-\textbf{p}_2)[1+F(\textbf{k},t)]\delta_{rr^{\prime}}\delta(\textbf{k}-\textbf{k}^{\prime})~, \notag
\end{align}
where the last equality is a consequence of the assumed spatial homogeneity of the system. More details of such kind of derivation in application to one-photon annihilation channel can be found in Ref.~\cite{Ref6}. 


For the electron-positron sector of EPP, the resulting non-Markovian type KE in the thermodynamic limit reads:
\begin{equation}\label{eq20}
\dot{f}(\textbf{p},t)=I(\textbf{p},t) + C^{(e)}(\textbf{p},t) + C^{(\gamma)}(\textbf{p},t) ,
\end{equation}
where the source term due to spontaneous pair production is defined by Eq. (\ref{eq12}), and the additionally arising collision integrals corresponding to the one-photon pair photoproduction/annihilation, and to the one-photon emission/absorption processes, respectively, are given by
\begin{subequations}\label{collisions}
\begin{eqnarray}\label{eq21}
C^{(e)}(\textbf{p},t)=&&\int \frac{d^3p_1}{(2\pi)^3}\int \frac{d^3k}{(2\pi)^3}\int^{t}dt^{\prime} K^{(e)}(\textbf{p},\textbf{p}_1,\textbf{k};t,t^{\prime})\notag\\
&&\{f(\textbf{p},t^{\prime}) f(\textbf{p}_1,t^{\prime}) [1+F(\textbf{k},t^{\prime})] - [1-f(\textbf{p},t^{\prime})][1-f(\textbf{p}_1,t^{\prime})]F(\textbf{k},t^{\prime}) \} ,
\end{eqnarray}
\begin{eqnarray}\label{eq22}
C^{(\gamma)}(\textbf{p},t)=&&\int \frac{d^3p_1}{(2\pi)^3}\int \frac{d^3k}{(2\pi)^3}\int^{t}dt^{\prime} K^{(\gamma)}(\textbf{p},\textbf{p}_1,\textbf{k};t,t^{\prime})\notag\\&&\{f(\textbf{p},t^{\prime})[1-f(\textbf{p}_1,t^{\prime})] [1+F(\textbf{k},t^{\prime})] - f(\textbf{p},t^{\prime})[1-f(\textbf{p}_1,t^{\prime})]F(\textbf{k},t^{\prime}) \} ,
\end{eqnarray}
\end{subequations}
in terms of the kernels
\begin{subequations}\label{kernels}
\begin{equation}\label{eq23}
K^{(e)}(\textbf{p},\textbf{p}_1,\textbf{k};t,t^{\prime})=\frac{(2\pi)^{3}e^2}{2k}\delta(\textbf{p}-\textbf{p}_1-\textbf{k}) Re\{[\overline{u}\upsilon]^{r+}_{\alpha\beta}(\textbf{p},\textbf{p}_1,\textbf{k};t) [\overline{u}\upsilon]^{r}_{\alpha\beta}(\textbf{p},\textbf{p}_1,-\textbf{k};t^{\prime})e^{-i\theta^{(+)}(\textbf{p},\textbf{p}_1,\textbf{k};t,t^{\prime})} \},
\end{equation}
\begin{equation}\label{eq24}
K^{(\gamma)}(\textbf{p},\textbf{p}_1,\textbf{k};t,t^{\prime})=\frac{(2\pi)^{3}e^2}{2k}\delta(\textbf{p}-\textbf{p}_1-\textbf{k}) Re\{[\overline{u}u]^{r+}_{\alpha\beta}(\textbf{p},\textbf{p}_1,\textbf{k};t) [\overline{u}u]^{r}_{\alpha\beta}(\textbf{p},\textbf{p}_1,-\textbf{k};t^{\prime})e^{-i\theta^{(-)}(\textbf{p},\textbf{p}_1,\textbf{k};t,t^{\prime})} \}
\end{equation}
\end{subequations}
constructed from the convolutions (\ref{eq7}) of the quasiparticle spinors (\ref{eq6}) and the phases
\begin{equation}\label{eq25}
\theta^{(\pm)}(\textbf{p},\textbf{p}_1,\textbf{k};t,t^{\prime})=\int_{t^{\prime}}^{t}d\tau[\omega(\textbf{p},\tau)\pm \omega(\textbf{p}_1,\tau)-k].
\end{equation}
Note that the resulting expressions in curly brackets in the collision integrals (\ref{eq21}), (\ref{eq22}) are written in a ``gain-minus-loss'' form, i.e. as differences of contributions from a direct and an inverse process, with the statistical weights being combined precisely in the way expected from Fermi/Bose statistics of (quasi-)electrons and (quasi-)positrons/photons. Furthermore, the delta-functions in the kernels (\ref{kernels}) reflect momentum conservation due to the assumed spatial homogeneity of the system, while the arising phases $\theta^{(\pm)}$ measure the amount of energy uncertainty in a process due to energy exchange with a time dependent semiclassical field. As easily seen, all these features are shared already by the initial equation (\ref{eq11}).

The photon sector KEs have the form similar to (\ref{eq20})
\begin{equation}\label{eq30}
\dot F(\textbf{k},t)=S^{(e)}(\textbf{k},t)+S^{(\gamma)}(\textbf{k},t),
\end{equation}
where the collision integrals for annihilation and emission channels read:
\begin{subequations}\label{collisions1}
\begin{eqnarray}\label{eq31}
S^{(e)}(\textbf{k},t)=&&\int \frac{d^3p_1}{(2\pi)^3}\int \frac{d^3p_2}{(2\pi)^3}\int^t dt^{\prime} K^{(e)}(\textbf{p}_1,\textbf{p}_2,\textbf{k};t,t^{\prime})\notag\\
&&\{ f(\textbf{p}_1,t^{\prime})f(\textbf{p}_2,t^{\prime})[1+F(\textbf{k},t^{\prime})] - [1-f(\textbf{p}_1,t^{\prime})][1-f(\textbf{p}_2,t^{\prime})]F(\textbf{k},t^{\prime}) \},
\end{eqnarray}
\begin{eqnarray}\label{eq32}
S^{(\gamma)}(\textbf{k},t)=&&\int \frac{d^3p_1}{(2\pi)^3}\int \frac{d^3p_2}{(2\pi)^3}\int^t dt^{\prime} K^{(\gamma)}(\textbf{p}_1,\textbf{p}_2,\textbf{k};t,t^{\prime})\notag\\
&& \{ f(\textbf{p}_1,t^{\prime})[1-f(\textbf{p}_2,t^{\prime})][1+F(\textbf{k},t^{\prime})] - f(\textbf{p}_1,t^{\prime})[1-f(\textbf{p}_2,t^{\prime})]F(\textbf{k},t^{\prime}) \} 
\end{eqnarray}
\end{subequations}
with the same kernels (\ref{eq23}), (\ref{eq24}). Thus, the kernels are universal for the fermion and photon sectors. Note that Eq.~(\ref{eq31}) was previously derived for the case $F\ll 1$ in Ref.~\cite{Ref6}.


The non-Markovian time $\tau=t-t'$, counted from the observation time $t$ the formation time of a process taken into account in a KE, describes both the slow quantum oscillations in the convolutions (vertex functions) contained in the kernels, and the rapid oscillations of the phases $\theta^{(\pm)}$. Thus the kernels (\ref{kernels}) can be further simplified by expanding the preexponential factors in Eqs.~(\ref{kernels}) in powers of $\tau$. For example, keeping just the zeroth order terms and using the sum rules that follow directly from the completeness of the set (\ref{eq6}) \cite{Ref6}
\begin{equation}\label{eq26}
u_{\alpha i}(\textbf{p})\,\overline{u}_{\alpha k}(\textbf{p})=\frac{(\hat{P}+m)_{i\,k}}{2\,\omega(\textbf{p})},\quad
\upsilon_{\alpha i}(\textbf{p})\,\overline{\upsilon}_{\alpha k}(\textbf{p})=\frac{(\hat{P}-m)_{i\,k}}{2\,\omega(\textbf{p})},
\end{equation}
where $\hat{P}=\gamma^\mu P_\mu$, we arrive at
\begin{equation}\label{eq28}
K^{(e,\gamma)}(\textbf{p},\textbf{p}_1,\textbf{k};t,t^{\prime})=\frac{(2\pi)^3e^2}{2k\,\omega(\textbf{p},T)\,\omega(\textbf{p}_1,T)}\delta(\textbf{p}-\textbf{p}_1-\textbf{k}) \Delta^{(\pm)}(\textbf{p},\textbf{p}_1,\textbf{k};t)\,\cos\theta^{(\pm)}(\textbf{p},\textbf{p}_1,\textbf{k};t,t^\prime),
\end{equation}
where (recall that sum over $r$ is implied)
\begin{equation}\label{eq29}
\Delta^{(\pm)}(\textbf{p},\textbf{p}_1,\textbf{k};t)=2(e^r(\textbf{k}) P)(e^r(-\textbf{k}) P_1)-[(PP_1)\pm m^2](e^r(\textbf{k})e^r(-\textbf{k}))
\end{equation}
and $(ab)=a_\mu b^\mu$. However, estimation of the precision and usefulness of such sort of approximation needs further investigation. Note that in Eqs.~(\ref{collisions}), (\ref{collisions1}) the distribution functions should be anyway kept unexpanded in $\tau$, as they use to suffer from high-frequency oscillations in a quasiparticle representation \cite{fedotovsmol2010}.




To close the system of equations one needs to add the Maxwell equation governing the plasma electric inner field \cite{Ref7,Ref14}
\begin{equation}\label{eq34}
\dot{E}_{in}(t)=-j_{cond}(t)-j_{pol}(t)=-j(t),
\end{equation}
where the conductivity and polarization current densities read:
\begin{subequations}
	\begin{equation}\label{eq35a}
	j_{cond}(t)=g\,e\int\frac{d^3p}{(2\pi)^3}\frac{P}{\omega(\textbf{p},t)}f(\textbf{p},t),
	\end{equation}
	\begin{equation}\label{eq35b}
	j_{pol}(t)=\frac{g\,e}{2}\int\frac{d^3p}{(2\pi)^3}\frac{\epsilon_{\perp}}{\omega(\textbf{p},t)}\left(u(\textbf{p},t)-\frac{e\dot{E}P}{4\omega^4(\textbf{p},t)}\right),
	\end{equation}
\end{subequations}
 $u(\textbf{p},t)$ is the polarization current function from Eq. (\ref{eq12.1}) and $g=4$ (product of fermion spin and charge degrees of freedom). The second term on the r.h.s. of (\ref{eq35b}) is a counter-term arising due to regularization.



\section{Summary}
\label{sec-4}

The closed system of equations (\ref{eq20}), (\ref{eq30}), and (\ref{eq34}) extends the well-known KE (\ref{eq11}) for EPP creation from vacuum \cite{Ref1,Ref2,Ref3,Ref4,Ref5} by including interaction with photons radiated by EPP. We take into account systematically both mechanisms of second order with respect to the parameter $E_q/E_c\ll1$: photon emission/absorption and the one photon pair photoproduction (along with the inverse process of one-photon annihilation). The KEs are closed with the Maxwell equation Eq.~(\ref{eq34}) determining the inner (plasma) field. While a nonperturbative kinetic description of $e^-e^+$-pair production from vacuum and a backreaction problem have been previously addressed, quantum radiation is included in such a model for the first time. 

The obtained equations can be used to study various aspects of EPP kinetics in a strong laser field, including radiation signatures from a strong field region (EPP diagnostics) \cite{Ridgers2013,Gonoskov1} and simulation of QED cascades \cite{Ref8,Ref9,Grismayer} beyond the locally constant field approximation. Since inverse processes are also included, a scenario of cascades saturation due to thermalization can be naturally addressed. The obtained equations may be useful in simulation of plural processes in heavy ions collisions, and also in condensed-matter physics (see, e.g., \cite{Ref13}). However, such particular applications are challenging and require further work, in particular on generalization to arbitrary polarization of the external field and on developing the corresponding efficient numerical solvers and will be addressed elsewhere.

The work was supported by the Tomsk State University Competitiveness Improvement Program, the MEPhI Academic Excellence Project (Contract No 02.a03.21.0005), the Foundation for the Advancement of Theoretical Physics ``BASIS'' (Grant 17-12-276-1) and the RFBR research project 17-02-00375a.

\end{document}